\documentclass[10pt, aps,prc,twocolumn,superscriptaddress,preprintnumbers,
amsmath, 
floatfix,
longbibliography,
nofootinbib
]{revtex4-1}
\usepackage[T1]{fontenc}
\usepackage{adjustbox}          
\usepackage[caption=false]{subfig}
\usepackage{url}
\usepackage{color}
\usepackage{float}
\usepackage[pdftex,colorlinks=true, linkcolor = blue, citecolor=blue,urlcolor=blue, bookmarksnumbered=true, bookmarksopen=true]{hyperref}
\usepackage{longtable}
\usepackage{amsfonts}
\usepackage{dsfont}
\usepackage{wrapfig,bm} 
\usepackage[normalem]{ulem}
\usepackage{MnSymbol}
\usepackage{float}

\newcommand{\beq}{\begin{equation}}
\newcommand{\eeq}{\end{equation}}
\newcommand{\bea}{\begin{eqnarray}}
\newcommand{\eea}{\end{eqnarray}}

\begin{document}

\title{Spatial orientation  of the fission fragment intrinsic spins and their correlations }

\author{Guillaume Scamps}
\affiliation{Department of Physics, University of Washington, Seattle, WA 98195--1560, USA}
\author{{Ibrahim Abdurrahman}} 
\affiliation{Theoretical Division, Los Alamos National Laboratory, Los Alamos, NM 87545, USA}   
\author{{Matthew Kafker}}
\affiliation{Department of Physics, University of Washington, Seattle, WA 98195--1560, USA}
\author{{Aurel Bulgac}}
\affiliation{Department of Physics, University of Washington, Seattle, WA 98195--1560, USA}
\author{{Ionel Stetcu}}
\affiliation{Theoretical Division, Los Alamos National Laboratory, Los Alamos, NM 87545, USA}   \date{\today}

\begin{abstract}
New experimental and theoretical results obtained in 2021
made it acutely clear that more than 80 years after the discovery of nuclear fission we do not understand the 
generation and dynamics of fission fragment (FF) intrinsic spins well, in particular 
their magnitudes, their spatial orientation, 
and their correlations. The magnitude and orientation of the primary FFs have a crucial role in defining 
the angular distribution and correlation between the emitted prompt neutrons, 
and subsequent emission of statistical (predominantly E1)
and stretched E2 $\gamma$-rays, and their correlations with the final 
fission fragments. Here we present detailed microscopic  
evaluations of the FF  intrinsic spins, for both even- and odd-mass FFs,
and of their spatial correlations. These point to a well-defined 3D FF intrinsic spin dynamics, 
characteristics absent in semi-phenomenological studies, due to the presence of the 
twisting spin modes, which artificially were suppressed in semi-phenomenological studies.

\end{abstract}  

\preprint{NT@UW-23-09,LA-UR-23-25884}

\maketitle   

 \vspace{0.5cm}
 
 The year 2021 started with the publication of a new and very accurate experimental 
 measurement of the fission fragments (FFs) intrinsic spins~\cite{Wilson:2021}, 
 significantly extending  the results of, almost 50-year-old, similar experiments~\cite{Wilhelmy:1972,Wolf:1976}. At the same time, an
 independent flurry of theoretical activity, based on phenomenological and microscopic models, were directed at studying various 
 properties of the FF intrinsic spins, which led to new insights into the mechanics of FF angular momenta formation and their 
 correlations~\cite{Vogt:2021,Bulgac:2021}.  Other theoretical 
studies followed ~\cite{Marevic:2021,Randrup:2021,Stetcu:2021,Bulgac:2022b,Bulgac:2022e,Randrup:2022,Randrup:2022a,Scamps:2022,Scamps:2023}, 
 and a very intense hybrid workshop was held to discuss the topic, attended by both theorists and experimentalists from all around the world, where many 
 new and old ideas were actively dissected~\cite{workshop:2022}.  As Lee Sobotka has recently discussed, in a talk at 
 the Nuclear Chemistry  Gordon conference in June 2023, we are now at a very unusual juncture in time, 
 when it is high time to address experimentally, {\it Fragment spin generation in Fission: 
 What we know, can't know, and should know}~\cite{Sobotka:2023}. 
 
 The case of spontaneous fission of $^{252}$Cf is perhaps the simplest and cleanest nucleus
 to consider in order to appreciate the complexity of what we need to 
 better understand the FF intrinsic spins, both theoretically and experimentally. In its ground state, $^{252}$Cf has the spin and parity $S_\pi=0^+$, and is a cold isolated 
 quantum system. After the FFs separate, both are highly excited, with the heavy FF (HFF) typically being cooler 
 than the light FF (LFF)~\cite{Bulgac:2016,Bulgac:2019c,Bulgac:2020}.  At the same time, the average intrinsic spin of the HFF is smaller compared to the LFF, as recently demonstrated in a first fully microscopic study on the FF intrinsic spin distributions~\cite{Bulgac:2021}. This was opposite to the prior consensus in literature, namely that the HFF 
 has a larger average intrinsic spin than the LFF~\cite{Wilhelmy:1972,Vogt:2009,Vogt:2013,Vogt:2021,Becker:2013,Litaize:2012}, 
 to cite a few representative studies.  This surprising result turned everything around making it clear that too much was taken for granted in modeling 
 fission dynamics and the decay properties of prompt FFs, which require a more detailed analysis.  Subsequent theoretical and phenomenological 
 studies incorporated this new aspect~\cite{Marevic:2021,Randrup:2021}. Recently, the relative angular momenta of the FFs was also investigated microscopically~\cite{Bulgac:2022b}.
 Conservation of the total angular momentum then requires that
 \begin{align}
 \hat{\bm S}_H +\hat{\bm S}_L + \hat{\bm \Lambda} = {\bm S}_0 \approx {\bm 0}, \label{equ:3}
 \end{align}
 where $\hat{\bm \Lambda} = \hat{\bm R}\times \hat{\bm P}$ is the relative orbital angular
 momentum perpendicular to the fission axis, $\hat {\bm R}, \hat{\bm P}$ are the 
the relative separation  between the FFs and their relative linear momentum respectively,  ${\bm S}_0$ is the compound's spin, and $\Lambda_z=0$. The above approximation is exact for $^{252}$Cf, and reasonable for the induced fission with low-energy incident neutrons on $^{235}$U, $^{239}$Pu targets. Now, a very important question arises:
are the FF intrinsic spins ${\bm S}_{H,L}$ also perpendicular to the fission axis? 
Clearly their sum $\hat{\bm S}_H +\hat{\bm S}_L$ is, in the case of 
spontaneous fission of $^{252}$Cf. This particular aspect is not yet resolved experimentally or theoretically~\cite{Randrup:2022}, and is related to the strong disagreements 
between TDDFT predictions~\cite{Bulgac:2022b,Bulgac:2022e}  and the phenomenological predictions of the
FREYA model~\cite{Randrup:2009,Vogt:2021,Randrup:2021,Randrup:2022}. 
The event-by-event orientation of the FF intrinsic spins has important 
consequences, as it will affect the direction of emission for prompt neutrons 
and is one of the most pressing questions experiment 
should now address~\cite{Sobotka:2023}. 
In the present microscopic study, we will specifically address this aspect and make a clear statement about
where the most advanced microscopic theory stands today, with a result starkly different from 
what the phenomenological model FREYA~\cite{Vogt:2021,Randrup:2021,Randrup:2022} predicts, which is the only other source of 
clear information available currently in literature.

The first indication that the angular distribution between the FF intrinsic spins is 
likely very different from previous models was reported in 
Refs.~\cite{Bulgac:2022b,Bulgac:2022e}. 
This  result was at odds with phenomenology implemented  
in FREYA~\cite{Vogt:2021,Randrup:2021,Randrup:2022}, where the angular distribution was almost uniform, while the 
microscopic results showed a clear non-uniformity. One major assumption adopted in FREYA, and which we demonstrate 
in the present study to be theoretically wrong, is that the FF intrinsic spins are perpendicular to the fission direction 
and as a result the twisting and tilting modes~\cite{Nix:1965,Moretto:1980,Moretto:1989}  
are artificially excluded from the fission dynamics.
At the time, when the theoretical study~\cite{Bulgac:2022b}  
was performed, an angular momentum projection on several angles was out of question. 
Today, two new technical  developments have made such a full study possible, see online supplement~\cite{supplement},
and presently 
one can easily evaluate the triple distribution 
$P(\Lambda,S_H,S_L)$ exactly, without any additional assumptions or approximations. 

The FF angular momentum projection is performed using well-known and established 
projection techniques~\cite{Varshalovich:1989,Ring:2004,Robledo:2009,Bertsch:2012,Bally:2021}, illustrated here for a specific FF
\begin{align}
&\hat P^{S}_{MK} = \frac{(2S+1)}{16\pi^2} \int \!\!d\Omega {\cal D}^{S*}_{MK} (\Omega) \,
e^{i\alpha \hat S_z} e^{i\beta \hat S_y} e^{i\gamma \hat S_z} , \label{eq:PJMK}\\
&P(S_F, K_F) = \langle \Psi | \hat P_{K_F K_F}^{S_F} | \Psi \rangle, \label{eq:Prob_JK}
\end{align}
with $\Omega=(\alpha,\beta,\gamma)$ representing a separate set of the three Euler angles  for each FF,
$|\Phi\rangle $ representing the many-body wave function, and $P(S_F,K_F)$ the probability distribution for 
either light or heavy intrinsic spin $S_F= S_L, S_H$ with projection $K_F$ on the fission direction. 
The angular momenta $\hat S_{x,y,z}$ are defined 
in a spatial region around a specific FF in its center-of-mass frame~\cite{Bulgac:2021}.  
$M$ and $K$ are the projections of the angular momentum $S$ in either the laboratory or body-frame.
Our goals are to evaluate $P(S_F, K_F) $ is  probability distribution for $K_F $ for each FF, 
and the triple angular momentum distribution 
\begin{align}
P(\Lambda,S_H,S_L)&= \sum_{k_H k_L} 
\langle \Psi | \hat P^{\Lambda}_{0,0} \hat P^{S_H}_{K_HK_H} \hat P^{S_L}_{K_LK_L} |\Psi\rangle. \label{eq:Triple_proj}
\end{align}
The triple distribution can be shown to be given exactly by the expression 
\begin{align}
&P(\Lambda,S_H,S_L) = \sum_{ K_H K_L K_H' K_L'}  (-1)^{K_H'-K_H+K_L'-K_L}\label{eq:Triple_proj2}\\ 
& \times C_{S_H,-K_H,S_L,-K_L}^{\Lambda,0}  
C_{S_H,-K_H',S_L,-K_L'}^{\Lambda,0}  
\langle \Psi | \hat P^{S_H}_{K_H K_H'} \hat P^{S_L}_{K_L K_L'} | \Psi \rangle. \nonumber
\end{align}
with $C_{j_1,m_1,j_2,m_2}^{J,M}$ the well-known Clebsch-Gordan coefficients, see online supplement~\cite{supplement}. 
This formula shares some common elements with a formula suggested by T. D\o{}ssing 
during the {\it Workshop of Fission Fragment Angular Momenta}~\cite{workshop:2022} 
and also discussed in Refs.~\cite{Bulgac:2022e,Scamps:2022,Scamps:2023}. 
The presence of the Clebsch-Gordan 
coefficients, which emerge naturally, ensure that the triangle constraint Eq.~\eqref{equ:3} is automatically satisfied. 
The numerical evaluations were performed using the 
LISE package~\cite{Shi:2020} to evolve the time-dependent density functional theory equations extended to superfluid systems and 
determine the many-body wave function $|\Phi\rangle$ used in Eq.~\eqref{eq:Triple_proj}.
In addition, at the end of the simulation, we performed a unitary transformation to the canonical quasi-particle states~\cite{Bulgac:2022c}, 
as they provide the most economic representation of a many-body wave function. 
For the evaluations of  the overlaps in Eq.~\eqref{eq:Triple_proj2}, which involves computing Pfaffians~\cite{Robledo:2009,Bertsch:2012}, we used 
the algorithm described in Ref.~\cite{Wimmer:2012}. 

 For each FF we define the angle between the ${\bm S}_{H,L}$ and 
 the fission axis, as well as the angle between 
 the two FF intrinsic spins~\cite{Bulgac:2021}
\begin{align}
&\!\!\!\!\! \cos \theta_F = \frac{K_F}{\sqrt{S_F(S_F+1)}},\,\text{where} \,  F = H, L \label{eq:theta}\\
&\!\!\!\!\! \varphi_{HL} = \arccos\left( \frac{ \Lambda (\Lambda +1) - S_H (S_H +1) - S_L(S_L+1) }{2 \sqrt{S_H(S_H + 1) S_L(S_L + 1)}  } \right). \label{eq:openang}
\end{align} 
Such angles can be defined if $S_{H,L}\ne 0$. With the triple distribution $P(\Lambda,S_H,S_L)$ 
one can straightforwardly evaluate the distributions $P(\theta_F)$ and $P(\varphi_{HL})$, which we will discuss now.  
 
 \begin{figure}
\centering
\includegraphics[width=.99\linewidth, keepaspectratio]{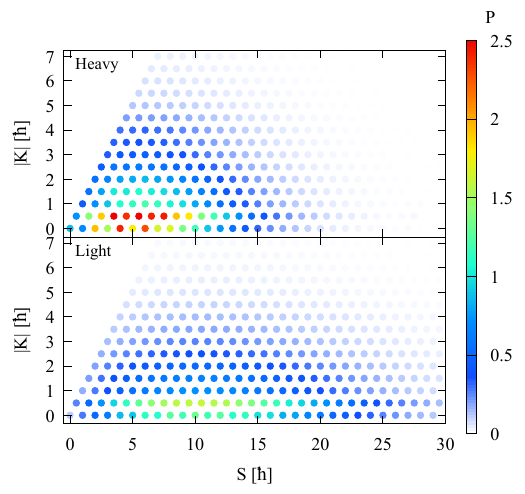} 
\caption{ Spin distribution normalized to 100 in the heavy (upper panel) and light (lower panel) fragment obtained with Eq.~\eqref{eq:Prob_JK}. 
This calculation is done for a $^{236}$U with the SeaLL1 functional. The states with the same integer $K$-values 
(unlike the half-integer $K$-values) for different $S_{H,L}$ 
are joined by a thin horizontal line for an easier visual identification. 
Since for both FFs $P(S,K)=P(S,-K)$ we show the distribution of $|K|$. } 
\label{fig:fig1}
\end{figure}  

\begin{figure}[h]
\centering
\includegraphics[width=.99\linewidth, keepaspectratio]{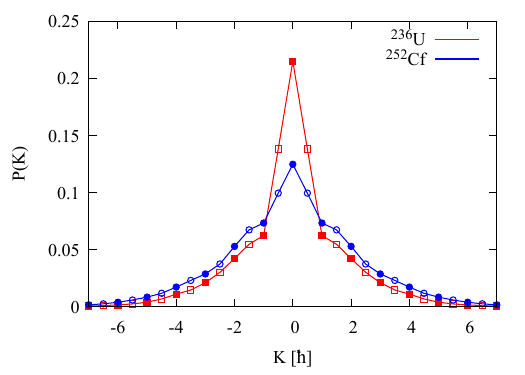} 
\includegraphics[width=.99\linewidth, keepaspectratio]{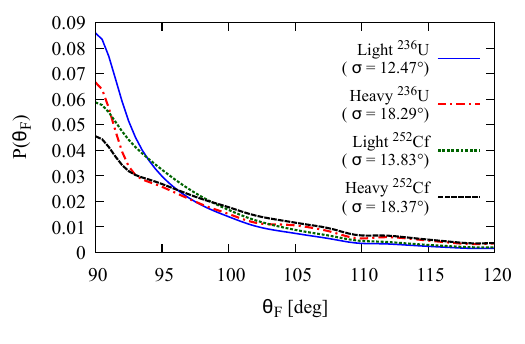} 
\caption{ Top: The distribution of the $K$ quantum number. Since $K_H+K_L=0$ the distributions for heavy and light FFs are identical. 
The integer $K$-values are shown with filled symbols and with empty symbols for half-integer $K$-values.
The bottom panel shows the distribution $P(\theta_F)$ computed using Eq.~\eqref{eq:theta}, convoluted with a Gaussian of width 2$^{\circ}$ 
and shown here only for angles $\theta_F\ge 90^\circ$, since $P(\theta_F)$ is symmetric with respect $\theta_F= 90^\circ$.  } 
\label{fig:fig2}
\end{figure}

The distributions for the projections of each FF spin on the fission axis are shown in Figs.~\ref{fig:fig1} and \ref{fig:fig2}.
Here one sees the first new aspects 
of each FF $K$-spin distribution and also, as one might have expected, the presence  
of both integer and half-integer spins. If $K$ is an integer, 
the corresponding FF is either an even-even (both $Z_F$ and $N_F$ are even) or an odd-odd nucleus  
(both $Z_F$ and  $N_F$ are odd). In the case of half-integer FF spins and $K$-values the corresponding $A_F$ is odd.
Additionally, the range of LFF intrinsic spins $S_L$ is wider than the range of the HFF intrinsic spins $S_H$, 
in agreement with the results reported  in Refs.~\cite{Bulgac:2021,Marevic:2021}.
Another noticeable aspect is that 
the probabilities to find even and odd mass FFs are almost equal. This is consistent with little or 
no odd-even staggering observed in experimental mass yields. Note that pre-neutron emission mass yields 
are corrected for neutron emission, correction that is subject to model dependence. 

The most remarkable aspect of these FF spin distributions shown in Fig.~\ref{fig:fig1} and of the data 
shown in Fig.~\ref{fig:fig2}, where the probability distributions 
\begin{align}
P(K)= P_H(K)=P_L(K)=\sum_{S_{L,H}}P(S_{L,H},K_{L,H})
\end{align}
are shown,
is the presence of non-vanishing values of the projection of each FF intrinsic spin on the fission direction, 
a incontrovertible confirmation of that fact that the twisting spin modes are active. 
This  feature is at stark odds 
with the almost 15 years old assumption made in FREYA~\cite{Randrup:2009,Vogt:2021,Randrup:2021,Randrup:2022} 
that the FF intrinsic spins are perpendicular to the fission axis and that the 
tilting and twisting modes of FF intrinsic spins are frozen and 
not active in the fission dynamics. The justification is based on an argumentation used
in the treatment of nucleon transfer in nuclear collisions~\cite{Randrup:1979,Randrup:1982}. 
This assumption played a key role in the claimed agreement~\cite{Randrup:2021} 
with the recent experimental results obtained by \textcite{Wilson:2021}. Since in FREYA
the FF intrinsic spins are treated classically there is no distinction between integer and half-integer FF 
intrinsic spins and no statement can be made about whether even-odd staggering effects are present in their predictions.

For both fissioning nuclei $^{252}$Cf and $^{236}$U one observes a very 
large peak corresponding to $|K| \le 1/2$, followed by quite long tails. In the cases of $^{252}$Cf and $^{236}$U, the
sum $P(-1/2)+P(0)+P(1/2) \approx 0.33$ and  $\approx 0.49$ respectively. Correspondingly, this implies that the probability to 
find a FF with $|K| \ge 1$ is 0.67 for $^{252}$Cf and $0.51$ for $^{236}$U. 
This is particularly important, as it points to the fact 
that the FF intrinsic spins are, with very large probability, not perpendicular to the fission axis. 
Instead, they are most likely to be found opposite to each other with respect to the fission axis 
(as  $K_H+K_L=0$). As a result, the plane defined by the triangle formed via the three angular 
momenta $\hat{\bm S}_H +\hat{\bm S}_L + \hat{\bm \Lambda} = {\bm 0}$ forms an angle $\theta_F$ 
significantly different than 90$^\circ$ with the fission axis for very large fraction ($\ge 1/2$) 
of fission events. The lower panel of Fig.~\ref{fig:fig2} 
reinforces this conclusion. From the results reported in Ref.~\cite{Bulgac:2021}, specifically
the expectation value of $K_F^2 \approx 1.6\ldots 4.4$  one obtains very similar values for $\theta_F$.

The wide range of active $K_{H,L}$ values is particularly important, as they 
control the so-called FF twisting degrees of freedom, whose role was ignored in the 
 FREYA model. In this respect 
one should also notice the role played by Coulomb re-orientation effects of the 
separated FFs~\cite{Scamps:2022,Scamps:2023},
which can lead to the increase of the FF intrinsic spins by 1-2 $\hbar$.  
Additionally, it contributes to the wriggling motion of the FFs, which otherwise 
is absent, since $K_H+K_L=0$ before scission in case of $^{252}$Cf(sf), 
unless the role of quantum fluctuations is taken into account 
explicitly~\cite{Bulgac:2019d,Bulgac:2020}.

\begin{figure}
\centering
\includegraphics[width=.99\linewidth, keepaspectratio]{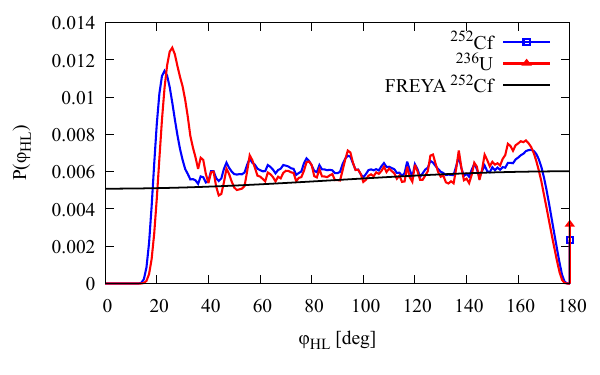} 
\caption{ The distribution $P(\varphi_{H,L})$, where $\varphi_{HL}$ is the opening angle between the FF intrinsic spins, 
 using Eqs.~\eqref{eq:openang} and \eqref{eq:Triple_proj2}. This distribution was obtained 
 after averaging with a $2^{\circ}$ wide Gaussian function.  The FF spin configuration at  
180$^\circ$ is shown as a separate point, as when seen with higher resolution this configuration 
is clearly separated from the configuration at $\varphi_{HL}<180^\circ$, 
but not as distinctly as the configurations close to $\varphi_{HL}=0^\circ$.
} 
\label{fig:fig3}
\end{figure} 

In Fig.~\ref{fig:fig3} we show the distribution of the opening angle between the FF intrinsic spins. 
This distribution, as reported in 
Refs.~\cite{Bulgac:2022b,Bulgac:2022e}, was completely at odds with the results arising from FREYA 
simulations~\cite{Vogt:2021,Randrup:2021,Randrup:2022}, and generated a lot of excitement and discussions 
at the fission workshop~\cite{workshop:2022}. As mentioned above, at the time, due to technical difficulties in
Refs.~\cite{Bulgac:2022b,Bulgac:2022e} we were not able to perform a full momentum projection and had to rely 
on the two-angle formulas. 
In the present study, this difficulty has been overcome and 
the exact triple angular momentum distribution is shown in Fig.~\ref{fig:fig3}. A comparison between 
the triple angular distributions $P(\Lambda,S_h,S_L)$ obtained in  Refs.~\cite{Bulgac:2022b,Bulgac:2022e} and 
the triple distribution evaluated in this study is presented in the online supplemental material~\cite{supplement}.
In the present microscopic treatment of the FF intrinsic spins, 
the distribution $P(\varphi_{HL})$ shows an almost uniform distribution in the interval 
$\varphi\in (40^\circ,160^\circ)$, with pronounced decays close to angles $0^\circ$ and $180^0$. 
The probability that the angle $\varphi_{HL}$ has values larger than $90^\circ$ is about 0.53, thus pointing to a slight preference for the bending over wriggling modes.

As intuited by T. D\o{}ssing during 
the workshop~\cite{workshop:2022}, there was a need for a Clebsch-Gordan coefficient to enforce the triangle constraint, 
similar, but not identical to the combination of Clebsch-Gordan coefficients in Eq.~\eqref{eq:Triple_proj2}.
The final results, shown in the upper panel of Fig.~\ref{fig:fig3}, are  closer to the  FREYA almost uniform 
predictions~\cite{Randrup:2009,Vogt:2021,Randrup:2021,Randrup:2022} for angles $(50^\circ,150^\circ)$.  
However, unlike FREYA predictions, the probabilities vanish at $\varphi_{HL} = 0^{\circ}$ and 180$^\circ$,
and obtain a prominent peak slightly above $\varphi_{HL}\approx 20^\circ$, and a smaller peak at $\approx 165^\circ$. 
These two peaks originate from the Clebsch-Gordan coefficients appearing Eq.~\eqref{eq:Triple_proj2} 
favoring angles $\varphi_{HL}$ close to $0^\circ$ 
and to a lesser extent $180^\circ$. 

The emerging lesson from our microscopic calculations 
is that the FF intrinsic spins dynamics is indeed of a 3D character and basically all FF intrinsic spin modes 
conjectured to be active  almost six decades ago~\cite{Nix:1965,Moretto:1980,Moretto:1989} are indeed present, 
see Figs.~\ref{fig:fig1} and \ref{fig:fig2}. 
These signals are clearest on the case of $^{252}$Cf(sf), in which case the angle formed 
by the plane defined by the FF intrinsic spins with the fission axis or the distribution $P(K)$ has 
very wide fluctuations.
The FF intrinsic spin dynamics is not restricted to the plane perpendicular to the fission axis, 
as in the classical treatment of the FF intrinsic spins of \textcite{Randrup:2009,Vogt:2021,Randrup:2021,Randrup:2022}. 
The FF collective twisting modes are clearly present, in agreement with the earlier 
conclusions in Ref.~\cite{Bulgac:2021}. 
The FF spin dynamics is also not fully unrestricted in 3D, as initially assumed in Ref.~\cite{Bulgac:2022b,Bulgac:2022e}.
The distribution of the angle between the  FF intrinsic spins reported initially in
Ref.~\cite{Bulgac:2022b}, with the details clarified and reported here 
in Figs.~\ref{fig:fig1}, \ref{fig:fig2} and \ref{fig:fig3}, 
are what is expected to either emerge or to be refuted in future envisioned experiments~\cite{Sobotka:2023,Randrup:2022a}. 
The fact that the FF intrinsic  spins are not  perpendicular to the fission axis before the emission of prompt neutrons 
and $\gamma$-rays, would likely lead to measurable  effects~\cite{Sobotka:2023,Randrup:2022a}. 

Often, either in discussions or in literature~\cite{Sobotka:2023}, one can find the statements that the pair-breaking mechanism 
can lead to 3D dynamics of the FF intrinsic spins. This aspect requires some clarifications. In microscopic   
studies~\cite{Bulgac:2016,Bulgac:2019d,Bulgac:2019c,Bulgac:2020,Bulgac:2022,Bulgac:2022b,
Magierski:2022} pairing is treated explicitly
and during the systems' evolution through
the saddle-to-scission descent, as well as in studies where the excitation energy of the initial compound nucleus was increased, 
the $nn$ and $pp$ short-range correlations (SRCs) never vanished, even though the excitation energy 
of the nuclear system corresponds to a high temperature, where a pairing condensate does not exist. 
Instead, only the phase of the pairing condensate is lost, true also in collisions of heavy-ions at rather 
large collision energies~\cite{Bulgac:2022,Magierski:2022,Bulgac:2022a,Bulgac:2022c}. 
SRCs between either proton or neutron pairs survive 
to rather large excitation energies, an aspect that should not be 
conflated with pair breaking. Loss of long-range order, manifested as the loss of phase coherence of the pairing condensate,
can be accompanied by the formation of new nucleon pairs with non-zero total spin.  
Nevertheless, the SRCs obviously survive in $L=0$.

The semi-phenomenological FREYA model~\cite{Randrup:2009,Vogt:2021,Randrup:2021,Randrup:2022}, 
which is based on a number of fitting parameters and assumptions, is the only model which so far leads to 
predictions, which might be tested in experiments. FREYA and the microscopic 
treatment of fission dynamics lead to starkly different predictions.  
This difference will be addressed by 
future experiments, which will be hopefully interpreted in fully assumptions and parameter-free 
theoretical treatments.  The microscopic framework adopted in this study is based on a nuclear 
energy density functional, which depends only 8 parameters (saturation density and energy of 
symmetric nuclear matter, spin-orbit and pairing couplings, proton charge, nuclear surface tension 
(related to the nucleon-nucleon interaction range), symmetry energy and to a less extent its 
density dependence~\cite{Shi:2018,Bulgac:2022f}, whose values are well-known for decades. 

As \textcite{Randrup:2022} stressed: \emph{
In view of the large differences between the model calculations of the spin-spin opening angle distribution, experimental
information on this observable is highly desirable as it could help to clarify the scission physics.} 
The assumption in FREYA the FF intrinsic (classical) spins are exactly perpendicular to the 
fission direction and that the twisting mode are inactive, is the origin of these large differences. As 
\textcite{Sobotka:2023} discussed, see slide 28, the angular distribution of FF stretched $\gamma$(E2) 
with respect to the fission direction is the ''smoking gun,'' which will discriminate between these two approaches and new experiments are planned to resolve discrepancies between old results~\cite{Wilhelmy:1972,Wolf:1976}.

 \vspace{0.5cm}

{\bf Acknowledgements} \\

We thank K. Godbey for a number of discussions of preliminary results.  We also thank L. Sobotka for his comments 
on  the manuscript.
The funding from the US DOE, Office of Science, Grant No. DE-FG02-97ER41014 and
also the support provided in part by NNSA cooperative Agreement
DE-NA0003841 is greatly appreciated. 
This research used resources of the Oak Ridge
Leadership Computing Facility, which is a U.S. DOE Office of Science
User Facility supported under Contract No. DE-AC05-00OR22725.
The work of I.A. and I.S. was supported by the U.S. Department of Energy through the Los
Alamos National Laboratory. The Los Alamos National
Laboratory is operated by Triad National Security, LLC,
for the National Nuclear Security Administration of the U.S.
Department of Energy Contract No. 89233218CNA000001.
I.A. and I.S. gratefully acknowledge partial support and computational
resources provided by the Advanced Simulation and
Computing (ASC) Program.


\providecommand{\selectlanguage}[1]{}
\renewcommand{\selectlanguage}[1]{}

\bibliography{local_fission.bib}

\end{document}


\title{Spin orientation of Fission Products Intrinsic Spins and their Correlations}

\author{G. Scamps$^{1}$, I. Abdurrahman$^{2}$, M. Kafker$^{1}$, A. Bulgac$^{1}$, and I. Stetcu$^{2}$}
\address{$^{1}$Department of Physics, University of Washington, Seattle, Washington 98195-1560, USA}
\address{$^{2}$Theoretical Division, Los Alamos National Laboratory, Los Alamos, New Mexico 87545, USA}

\begin{abstract} 
Here we provide some pertinent details to the main text.
\end{abstract}
\maketitle

\section{Derivation of the Eq. (5) in the main text} 

 The distribution of fission fragments (FFs) intrinsic spins $S_H,S_L,$ and of the orbital angular momentum ${\bf \Lambda} = {\bf R}\times {\bf P}$,
 where ${\bf R}$ is the FF relative separation and ${\bf P}$ their relative linear momentum, is obtained using the triple projection~\cite{Bulgac:2022b}, 
\begin{align}
P(\Lambda,\Lambda_z,S_H,S_L)&= 
\sum_{K_H K_L} \langle \Psi | \hat P^{\Lambda }_{\Lambda_z\Lambda_z} \hat P^{H S_H}_{K_HK_H} \hat P^{L S_L}_{K_LK_L} |\Psi\rangle,\label{eq:Triple_proj}
\end{align}
where the symmetry projectors are defined as~\cite{Ring:2004,Varshalovich:1988,Bally:2021}
\begin{align}
& \hat P^{F J_F}_{K_F K_F'}= \frac{(2J+1)}{16\pi^2} \iiint \hat R_F (\Omega_F) {\cal D}^{J_F*}_{K_FK_F'} (\Omega_F) d\Omega_F, \label{eq:PJMK} \\
&   \hat R_H (\Omega_H)  \hat R_L (\Omega_L)=\hat R_L (\Omega_L)  \hat R_H (\Omega_H),\\
& \hat P^{\Lambda}_{\Lambda_z \Lambda_z}= \frac{(2\Lambda+1)}{8\pi^2} 
\iiint \hat R (\Omega) {\cal D}^{\Lambda*}_{\Lambda_z,\Lambda_z} (\Omega) d\Omega, \label{eq:PL}\\
&\hat R (\Omega) = e^{i\alpha \hat J_z} e^{i\beta \hat J_y} e^{i\gamma \hat J_z},
\end{align}
and $\hat R (\Omega)$  rotates the fission direction and  $\hat R_F (\Omega_F)$ rotate the FFs respectively. 
The rotation of each FF should be performed with respect to their center of mass and in their moving framework.
Since the operator 
\begin{align}
\hat R(\Omega)\hat R_H(\Omega) \hat R_L(\Omega)
\end{align} 
rotates the entire system in Eq.~\eqref{eq:Triple_proj}  $\hat R(\Omega)$ can be replaced by $\hat R_H(-\Omega) \hat R_L(-\Omega)$, 
which has the effect of rotating simultaneously both FFs. Note that $\Lambda_z = 0$ in the intrinsic system of the fissioning nucleus. 
Introducing for each FF $F = H, L$, 
\begin{align} 
& \hat R_F (\Omega_F')    =  \hat R_F (-\Omega)  \hat R_F (\Omega_F) \\
& {\cal D}^{J_F*}_{M_F,K_F} (\Omega_F) = \sum_{M_F'} {\cal D}^{J_F'}_{M_F,M_F'} (- \Omega)  {\cal D}^{J_F'*}_{K_F,M_F'} (\Omega_F') 
\end{align}
 we obtain,
\begin{align}
P(\Lambda,S_H,S_L) &= \sum_{M_H,M_L} \frac{ (2\Lambda+1)  (S_H+1/2) (S_L+1/2)}{(8 \pi^2 )^3} 
\iiint d\Omega \iiint d\Omega_H'  \iiint d\Omega_L'   \langle \Psi |  \hat R_{H} (\Omega_H')   \hat R_{L} (\Omega_L')   | \Psi \rangle \nonumber \\
& \times {\cal D}^{\Lambda*}_{0,0} (\Omega)   {\cal D}^{S_H}_{0,-M_H} (\Omega) {\cal D}^{S_H*}_{0,M_H} (\Omega_H')  
{\cal D}^{S_L}_{0,-M_L} (\Omega)  {\cal D}^{S_L*}_{0,M_L} (\Omega_L') (-1)^{M_H+M_L} 
\end{align}
and we used the property Eq.~(5), page 97 in Ref.~\cite{Varshalovich:1988}),
\begin{align}
\iiint  d\Omega  {\cal D}^{J*}_{M,K} (\Omega)    {\cal D}^{J_1}_{M_1,K_1} (\Omega)  {\cal D}^{J_2}_{M_2,K_2} (\Omega) = \frac{8 \pi^2}{2J+1}
C^{JM}_{J_1M_1J_2M_2} C^{JM}_{J_1K_1J_2K_2}
\end{align}
where $C_{J_1,M_1,J_2,M_2}^{JM}$ is a Clebsch-Gordan coefficient,
\begin{align}
 {\cal D}^{J}_{M,K} (- \Omega) = (-1)^{K-M} {\cal D}^{J}_{-M,-K} (\Omega).
\end{align}
The expression \eqref{eq:Triple_proj} can be simplified exactly to the following projection formula,
\begin{align}
P(\Lambda,S_H,S_L) &= \sum_{ K K' }  C_{S_H,-K,S_L,K}^{\Lambda,0}  C_{S_H,-K',S_L,K'}^{\Lambda,0}  
\langle \Psi | \hat P^{H S_H}_{K K'} \hat P^{L S_L}_{-K -K'} | \Psi \rangle. \label{eq:Triple_proj2}
\end{align}
The use of Eq.~\eqref{eq:Triple_proj2} requires a projection on six Euler angles, which as far as we are aware of has never been attempted in practice. 
In the case when $\Lambda_z = K_H+K_L=0$ the number of needed Euler angles can be reduced to four, a number which is quite large and rarely used in practice.

\begin{figure}[t]
\centering
\includegraphics[width=0.45\columnwidth]{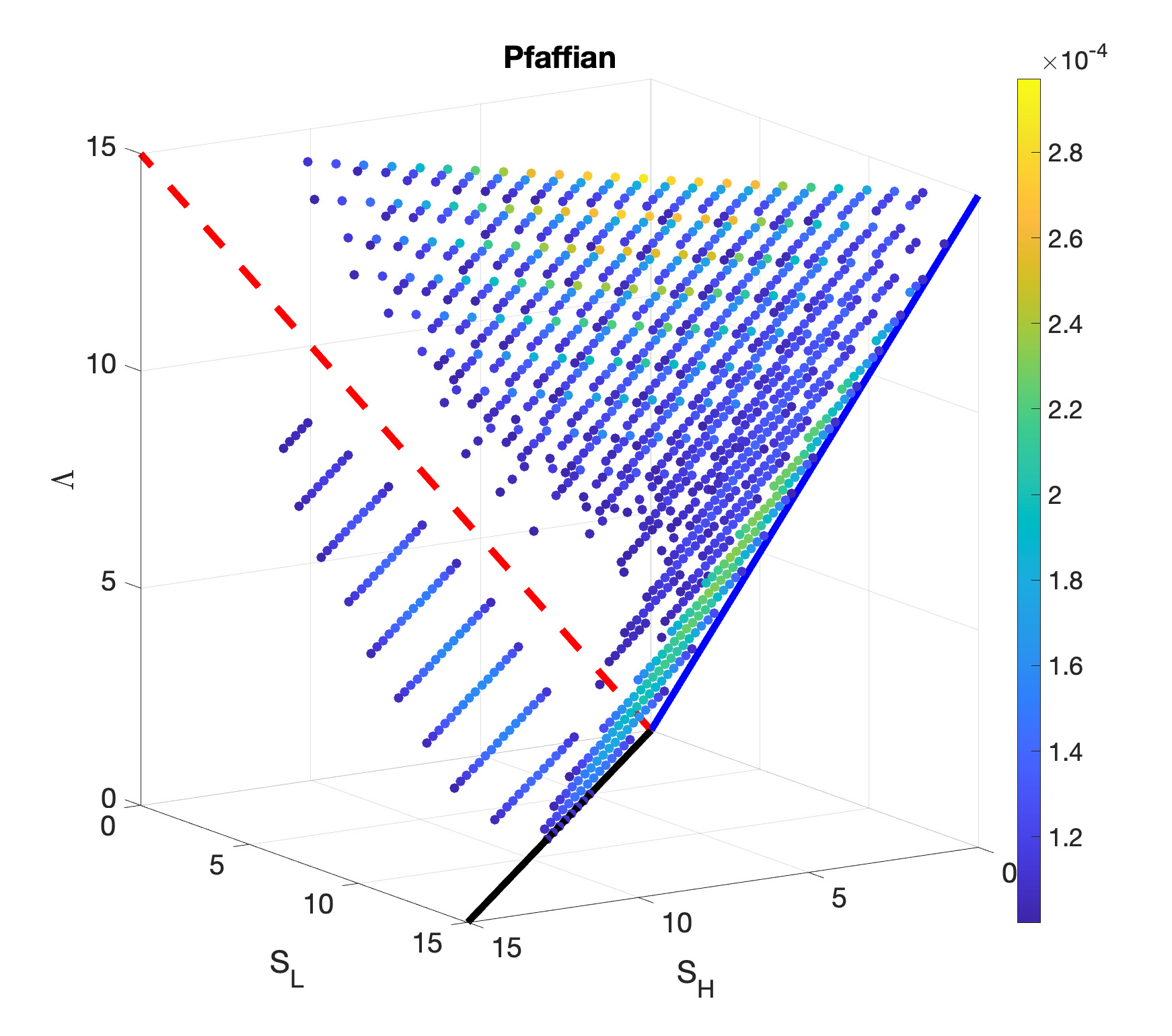}
\includegraphics[width=0.45\columnwidth]{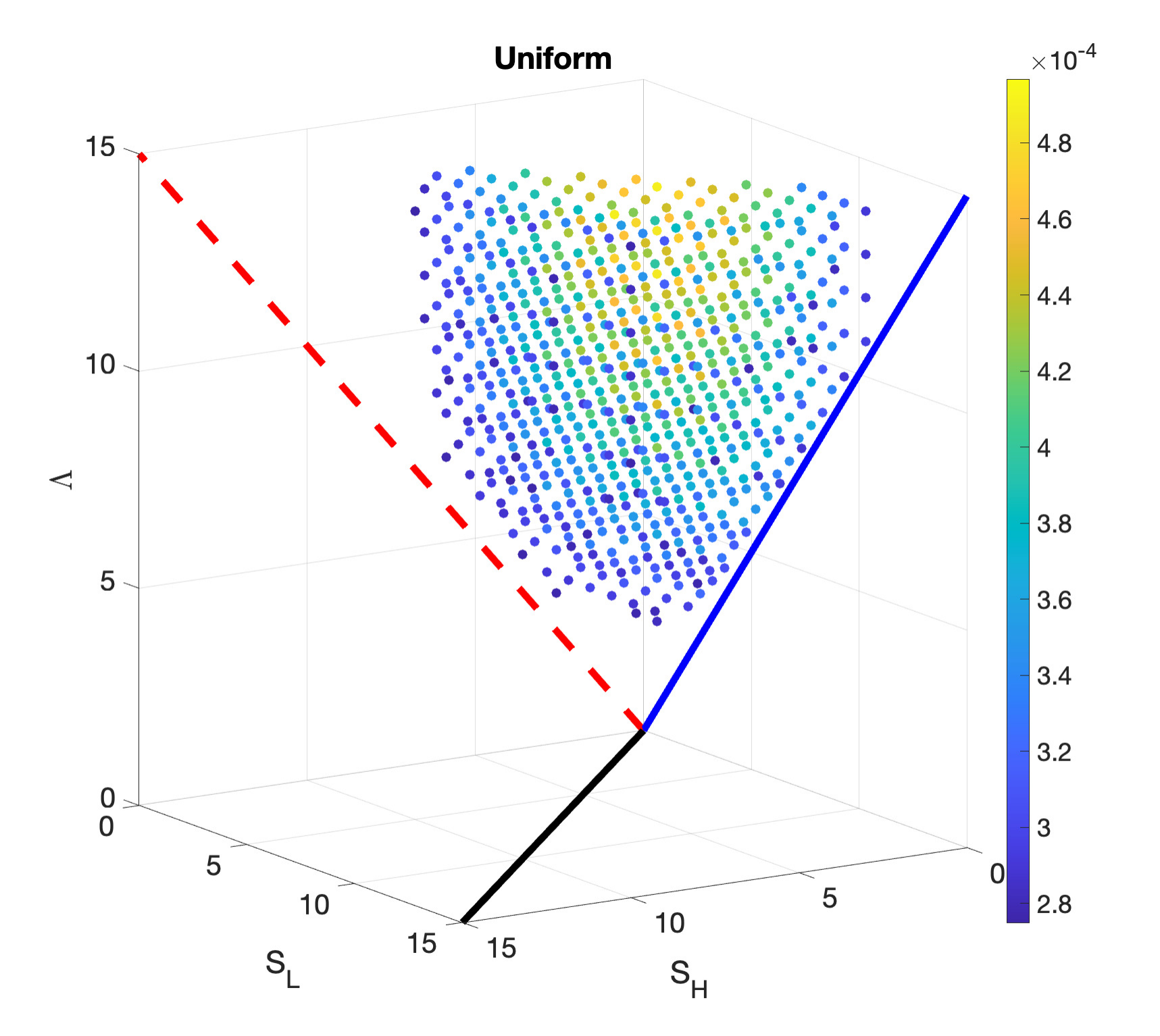}
\caption{ The left panel is the the distribution $P(\Lambda,S_H,S_L)$ obtained with Eq.~\eqref{eq:Triple_proj2} evaluated using the 
Pfaffian method described in Ref.~\cite{Bertsch:2012}, while in the right panel is the distribution obtained 
in Refs.~\cite{Bulgac:2022b,Bulgac:2022e}. 
The black, blue, and red-dashed lines correspond to one of the angular momenta $S_H, S_L, \Lambda$ vanishing and they meet at the origin
 $S_H=S_L= \Lambda=0$. Only the angular momenta triples contributing to the upper 30\% of the probability distribution $P(\Lambda,S_H,S_L)$ are shown. }
 \label{fig:spat}
\end{figure}

\section{Quasiparticle, canonical wave functions,  and Distributions $P(\Lambda,S_H,S_L)$}

We performed simulations in a box $N_x\times N_y\times N_z = 30\times 30\times 60$ with a lattice constant 1 fm, see Ref.~ \cite{Shi:2020}, with which we evolved 
$16\times30^2\times60= 864,000$ coupled nonlinear complex partial differential equations in 3D+time.  
The total number of quasiparticle wave functions being too large to perform 
the evaluations of the overlaps in Eqs.~(\ref{eq:PJMK}, \ref{eq:Triple_proj2}), we introduced canonical wave functions as described in Ref.~\cite{Bulgac:2022c}, which reduced 
significantly the dimension of the overlap matrix to be evaluated with the Pfaffian method described in Ref.~\cite{Bertsch:2012}.  
While the use of quasiparticle wave functions is  typical in treatment of nuclear systems with pairing, since the generalized density matrix commutes with the 
generalized quasiparticle Hamiltonian and one can define then in a natural manner the quasiparticle energies~\cite{Ring:2004}, 
the canonical wave functions are not eigenfunctions of the generalized pairing Hamiltonian and are very rarely used in practice. 
Since the canonical occupation probabilities are not conserved during time evolution, the set of canonical wave functions has to be introduced 
only in the final state, see Ref.~\cite{Bulgac:2022c}.

In Refs.~\cite{Bulgac:2022b,Bulgac:2022e}, we were forced to use a very large set of quasiparticle 
wave functions in order to evaluate the FF intrinsic spins, since the technology to determine the 
canonical wave functions and many of their properties described in Ref.~\cite{Bulgac:2022c} were not known. 
That prevented us from performing full symmetry projection and 
we have adopted at the time a reasonable approximation described here as well. 
We evaluated the triple distribution 
\begin{align}
\tilde{P}_0(\Lambda,S^L,S^H) &= \int_{-\pi/2}^{\pi/2}\frac{d\beta_0}{\pi}
\int_{-\pi/2}^{\pi/2}\frac{d\beta^L}{\pi}
\int_{-\pi/2}^{\pi/2}\frac{d\beta^H}{\pi}
e^{i\beta_0\Lambda + i\beta^LS^L +i\beta^HS^H} \nonumber \\
&\times \langle \Psi | 
e^{-i\beta_0(\hat{J}^L_x+J^H_x) -i\beta^L\hat{J}^L_x-i\beta^H\hat{J}^H_x}
|\Psi\rangle ,    
\end{align}
which technically depends on two angles only as an approximation to the projection 
\begin{align}
P(\Lambda,S_L,S_H)= \langle \Psi | \hat P^{\Lambda }_{0,0} \hat P^{H S_H}_{0,0} \hat P^{L S_L}_{0,0} |\Psi\rangle,
\end{align}
which unfortunately, as explained in Refs.~\cite{Bulgac:2022b,Bulgac:2022e} 
does not enforce the triangle constraint.
\begin{align}
\Delta = \Theta (\Lambda \ge |S_l-S_h|)\Theta (\Lambda \le S_L+S_H).
\end{align} 
For that reason we have used the approximate projector
\begin{align}
\tilde{P}(\Lambda, S_L,S_H) = {\cal N}
\tilde{P}_0(\Lambda,S_L,S_H)\Theta (\Lambda \ge |S_L-S_H|)\Theta (\Lambda \le S_L+S_H)    \label{eq:2}
\end{align}
with the appropriate normalization ${\cal N}$. The differences between 
the exact triple distribution with Eq.~\eqref{eq:Triple_proj2} and
the approximate triple distribution 
with Eq.~\eqref{eq:2} are illustrated in the 3D profiles in Fig.~\ref{fig:spat}.

At the level of one spin and two spins distributions these triple 
distributions lead to very similar results however. 
From the triple distributions $P(\Lambda,S_H,S_L)$ or $\tilde{P}(\Lambda,S_H,S_L)$ 
one can determine the single and double FF intrinsic spin distributions
\begin{align}
&P_1(S_L)= \sum_{S_H,\Lambda} P(\Lambda,S_H,S_L), \quad P_1(S_H)= \sum_{S_L,\Lambda} P(\Lambda,S_H,S_L),\\
& P_2(S_H,S_L) = \sum_\Lambda   P(\Lambda,S_H,S_L).
\end{align}
Our results show that 
\begin{align} 
P_1(S_F)\approx (2S_F+1)\exp \left [ -\frac{S_F^2}{2\sigma_F^2}\right ] ,\label{eq:P1}\\
\end{align}
a result derived in a Fermi gas model, apart from odd-even effects, see Refs.~ \cite{Bethe:1936,Ericson:1960}, 
and that separately for half-integer and integer spins
\begin{align}
P_2(S_H,S_L) \approx P_1(S_H)P_1(S_L),\label{eq:P2}
\end{align}
also observed experimentally by \textcite{Wilson:2021} and theoretically in
Refs.~\cite{Vogt:2021,Randrup:2021,Bulgac:2022e}.

In the present study, we were able to perform angular momentum projection with four Euler angles
(instead of formally six Euler angles) after introducing 
the set of canonical wave functions and taking into account the symmetry of the FFs.
The use of the approximate two-angle projector Eq.~\eqref{eq:2} is not necessary anymore. 
As a result, we were able to evaluate the intrinsic spins and their correlations for both 
even and odd mass FFs, as demonstrated in Fig. 1 in the main text and also evaluated 
the distribution of the projection of the 
FF intrinsic spins on the fission direction, see Figs. 1 and 2 in the main text. 

Due to these two technical developments, we were able to  prove 
that in fission dynamics all possible FF spin modes (conjectured to be allowed since
1965~\cite{Nix:1965,Moretto:1980,Moretto:1989},
but never incontrovertibly shown to exist) are excited at scission and beyond, 
where the nuclei are already hot. The possible FF spin modes are two 
bending and two wriggling modes 
and one twisting and another tilting~\cite{Nix:1965}.  
The tilting mode is excited only if the total spin of the compound nucleus ${\bf S_0} > 0$.
Surprisingly, the twisting mode is present 
with a probability larger than 0.5 for both  induced 
fission of $^{236}$U and spontaneous fission of $^{252}$Cf. 
This aspect is in stark contrast with the 
conjecture in Ref.~\cite{Nix:1965},  adopted subsequently in numerous data analyses, and the 
predictions of the phenomenological model introduced almost 15 years ago
FREYA~\cite{Randrup:2009,Vogt:2021,Randrup:2021,Randrup:2022}, 
where the twisting mode is deemed irrelevant and thus 
suppressed.  One should add that the generation of each single point 
in Figs. 1-2 in the main text require a separate angular 
momentum projection for each set of values $ S_F, K_F$ 
for the single intrinsic spin distribution  $\hat{P}^{F S_F}_{K_FK_F}$, see 
in Eq.~\eqref{eq:PJMK}, and similarly for each set of values $S_H, S_L, K, K', -K, -K'$ 
for the triple angular momentum distribution  
$P(\Lambda,S_H,S_L)$, see Eq.~\eqref{eq:Triple_proj2}, 
therefore leading to many hundreds of angular momentum projections performed in this study.

\bibliography{local_fission1.bib}